# μRANIA-V: an innovative solution for neutron detection in homeland security

R. Farinelli,[a,*] I. Balossino,[a,b] G. Bencivenni,[c] G. Cibinetto,[a] G. Felici,[c] S. Fiore,[d] I. Garzia,[a] M. Gatta,[c] M. Giovannetti,[c], R. Hall-Wilton,[e] C. C. Lai,[e] L. Lavezzi,[f] G. Mezzadri,[a,b] G. Morello,[c] E. Paoletti,[c] G. Papalino,[c] A. Pietropaolo,[d] M. Pillon,[d] M. Poli Lener,[c] L. Robinson,[e] M. Scodeggio,[a] P.O. Svensson.[e]

[a] INFN-Ferrara, via Saragat 1, 44122 Ferrara, Italy
[b] Institute of High Energy Physics, Chinese Academy of Sciences, 19B Yuquan Road, 100049 Beijing, China
[c] INFN-National Laboratory of Frascati, via Enrico Fermi 54, 00044 Frascati (Roma) Italy
[d] ENEA, via Enrico Fermi 45, 00044 Frascati (Roma), Italy
[e] Detector Group, European Spallation Source ERIC (ESS) Box 176, SE-221 00 Lund, Sweden
[f] INFN-Torino, via P.Giuria 1, 10125 Torino, Italy
[*] Corresponding author

rfarinelli@fe.infn.it

*Abstract*—Detection of neutrons is becoming of the utmost importance, especially in the studies of radioactive waste and in homeland security applications. The crisis of $^3$He availability has required the development of innovative techniques. One solution is to develop light gas detectors for neutron counting to be used as portals for ports and airports. The neutron is converted on the Boron-coated cathode, releasing a charged particle, whose passage can be identified by the gas detector. While several technologies have been deployed in the past, the project **μRANIA-V (μRwell Advanced Neutron Identification Apparatus) aims to detect thermal neutrons by means of the μRwell technology, an innovative gas detector. The goal is to produce tiles to operate as portals in homeland security or for radioactive waste management. The technological transfer towards the industry has started, thus the production can be cost-effective also owing to a construction process relatively easier compared to similar apparatus. By reading directly the signals from the amplification stage, the neutrons can be counted with simplified electronics further reducing the total cost. In this paper, the project will be described, with details on the μRwell technology and on the neutron counting, on the test beam performed, and on the future plans.**

*Keywords* — MPGD, neutron detection, radioactive waste, homeland security.

## I. Introduction

Neutrons are subatomic particles with neutral electric charge and a mass slightly greater than that of a proton. Neutrons have high penetration power and together with neutron detectors are used to probe the structure and the motion of the matter in a complementary way to X-ray imaging [1]. Neutrons play an important role with radioactive materials: there a neutron detector can be used as a monitor for those materials or even as a portal monitor for homeland security.

Neutrons are involved in nuclear transmutations such as the nuclear fission and free neutrons cause ionizing radiation with biological hazard, depending on dose. On Earth, a natural neutron background flux exists caused by cosmic ray showers and natural radioactivity. Free neutrons can be produced by neutron sources like neutron generators, research reactors and spallation sources [2-3]. In literature, neutron detection is categorized according to the nuclear processes, mainly neutron capture and elastic scattering (not threatened in this paper). The neutron detection happens by means of the neutron caption reactions into electrical signals through particles and energy released. Nuclides such as $^3$He, $^6$Li, $^{10}$B and other heaviers like $^{235}$U have a high neutron capture cross section and a larger probability of absorbing a neutron, as reported in Table 1.

| Reaction | Q-value [MeV] | Cross section for thermal neutrons [barns] |
|---|---|---|
| $^{10}$B + n → $^7$Li + α | 2.31 | 3840 |
| $^6$Li + n → $^3$H + α | 4.78 | 940 |
| $^3$He + n + $^3$H + p | 0.754 | 5330 |
| $^{235}$U + n + X + Y | ~200 | 575 |

Table 1. Reactions of interest involving neutron capture, the released energy (Q-value) and their cross-section for thermal neutrons are shown, where *n* represents a neutron, *p* a proton, α an alpha particle and *X/Y* the fission fragments.

The cross section of the neutron capture strictly depends on the neutron energy[4]: lower cross sections are measured for fast neutrons and higher values are observed for the slow ones. Calorimeters are used in the high energy regime where the neutron is totally absorbed and the deposited energy is measured; scattering with protons is exploited in a moderate energy regime; while in the low energy range mediums with large capture cross-section are used. A well known converter used for neutron detection is the $^3$He due to its large cross-section (5330 at 25 meV) but in the latest years a shortage of this element increased the $^3$He cost with a large impact on the neutron detector production and their application in the fields of interest. $^3$He is not extracted by natural resources but it is of industrial origin, precisely stemming from the decay of Tritium. In the Cold War a large amount of Tritium was accumulated but at present days the increasing interest in neutron detection and the reduced amount of $^3$He in the storage addressed its shortage. A call for alternative solutions triggered the URANIA-V project[5].

## II. Boron-coated gaseous detectors

Thanks to the neutron capture process, the neutron detection can be achieved with gaseous detectors through a thin surface of a converter: the reaction products are generated on the converter. Once they interact with the gas, the ionization path is amplified by a Micro Pattern Gaseous Detector (MPGD). The MPGD technology exploits the modern photolithography and thin layer polyimide depositions to develop a new design to amplify the ionization charge and to readout the signal. The granularity of the detector improves the intrinsic rate capability and the spatial resolution below to 50-100 μm. Triple-GEM, MicroMegas and μRWELL are examples of MPDG of success used for the particle detection; within the URANIA-V project the neutron detection is performed with a μRWELL-based detector.

The μRWELL is an innovative MPGD that combines the best features of the GEM amplification stage with a resistive layer such as in the MicroMegas detector. This led to a compact, spark-protected, resistive MPGD with a single amplification stage. The μRWELL detector is composed of two elements, cathode and the readout-PCB, to simplify its production process and to reduce the gap with the technological transfer.

The cathode is based on a Cu layer on a glass epoxy plate. Applications for neutron detection need a $B_4C$ deposition of a few μm to convert the thermal neutron into α or $^7$Li particles. The readout-PCB combines the anode, the amplification stage and the segmented readout. A WELL patterned Apical foil with a thin copper layer is used to amplify the ionization signal up to a gain of $10^4$. A Diamond-Like-Carbon (DLC) layer of resistivity of about 80-100 MΩ/☐ is used to quench the sparks and it does not affect the detector gain up to 1-10 MHz/cm$^2$. The detector reaches an efficiency of 98%, a spatial resolution of about 50 μm and a time resolution of 5 ns. Those performances match with the purpose of the project to develop a nuclear waste monitor, a homeland security portal and a neutron imaging detector.

## III. Neutron converter designs

The chosen converter in the URANIA project is the Boron. A $B_4C$ deposition on the cathode with sputtering technique was studied together with the colleagues of the ESS Coating Workshop in Linkoping, Sweden. A *planar converter* layer was used for the first design of a neutron detector together with the μRWELL detector. The $B_4C$ thickness is the first parameter to optimize the neutron detection efficiency: large amounts of Boron increase the conversion efficiency but as the thickness increases too much, then the reaction production (α and $^7$Li) can not enter into the gas volume, reducing the detection efficiency. The detection efficiency depends also on the neutron impinging angle with respect to the $B_4C$ surface: as long as the angle goes from a orthogonal to parallel then the neutron path inside the Boron layer is larger and it increases the conversion efficiency. The planar cathode design is the simpler one to be assembled. The performance of this converter was studied through simulation and experimental measurements as described in the next sections.

A second cathode design was studied to increase the detection efficiency of the detector. The neutron detection depends on their impinging angle then a *grooved cathode* was designed to develop a compact histrument with one cathode and a readout-PCB. Machining operation shapes the cathode with peaks of about 2 mm and a spacing of 0.5-1.5 mm then a $B_4C$ sputtering procedure is performed.

In addition, a third cathode design is under study to improve the detection efficiency. By means of a *mesh* between the cathode and the readout-PCB it is possible to introduce another electrodes within the active volume and with a proper electrical configuration it is possible to collect the primary ionization electron through the mesh up to the readout while the mesh itself acts as another neutron conversion stage thanks to a $B_4C$ layer sputtered on it. A mesh with a wire of 53 μm and a pitch of 127 μm was chosen for the measurements to prove the concept.

The second and the third designs are more complex because they have to take into account the geometry for the mesh or the grooved cathode to optimize the neutron conversion and the electron transparency of the detector to collect as much signal as possible. The three detector designs are summarized in Fig. 1.

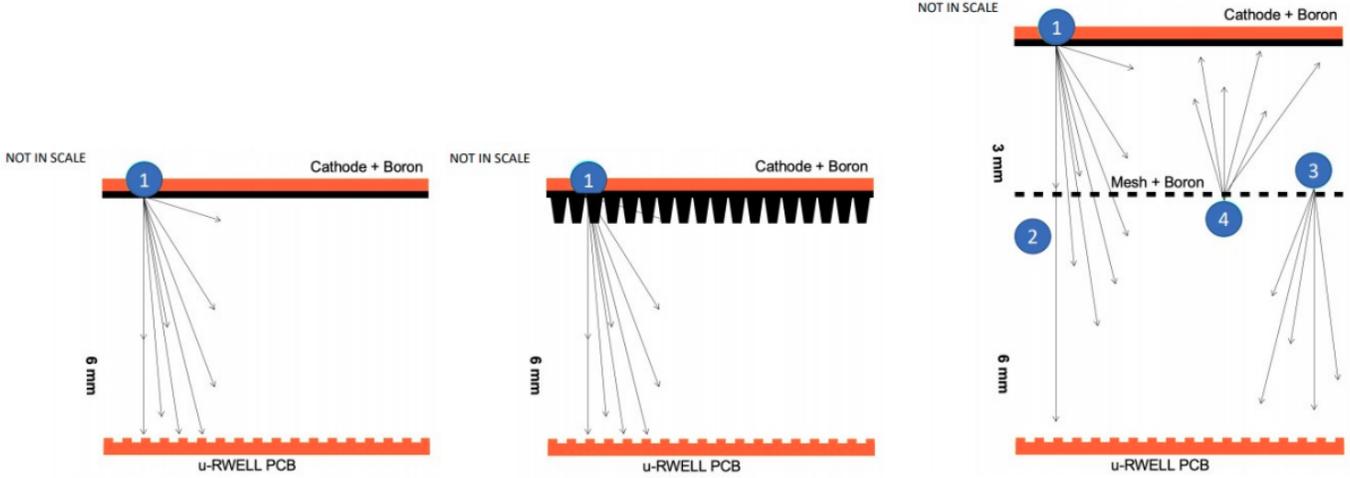

Fig. 1. Summary of the three detector designs under study in the URANIA-V project. From the left ot the right: planar cathode, grooved cathode and planar cathode with a mesh. Cathode and mesh are boron-coated. The numbers from 1 to 4 summarize the different processes.

## IV. SIMULATIONS

The first tool to approach neutron detection is simulation. Nowadays, computer software can accurately simulate the particles interaction through gas and matter, i.e. neutrons, alpha, lithium and electrons. The main simulators used are Geant4 and Garfield++ [6,7]. The first one is a toolkit for the simulation of the passage of particles through matter. The second is a toolkit for the detailed simulation of particle detectors based on ionisation measurement in gases and semiconductors. These tools allow us to measure the interaction probability of the neutron with the Boron layer and the other detector material that may participate in the interactions with the neutrons. Once the neutron capture happens, then the reaction products are studied to determine their path in the gas mixture, the active volume of the μRWELL detector. Once the primary ionization takes place, the electrons have to be properly collected on the readout plane: simulations can provide detailed knowledge of the processes. The first study is focussed on the optimization of the $B_4C$ thickness in the conversion layer sputtered on a *planar cathode*: by means of thermal neutron impinging orthogonally on the layer, several thickness is studied and the amount of α and $^7Li$ particles reaching the gas volume is measured, as shown in Fig. 2. Energy distributions are described by the two reactions involved in (1).

$$n + {}^{10}B \rightarrow {}^{11}B^* \rightarrow \alpha + {}^7Li \text{ (6\%)}$$
$$n + {}^{10}B \rightarrow {}^{11}B^* \rightarrow \alpha + {}^7Li + 477 \text{ keV } \gamma (94\%) \quad (1)$$

No processes observe α and $^7Li$ at the same time because their emission is almost back to back, therefore only one enters the gas volume. The production of $\gamma$ ray particles through the detector is much higher with respect to the charged particles but their impact is negligible up to detector gain of $10^4$.

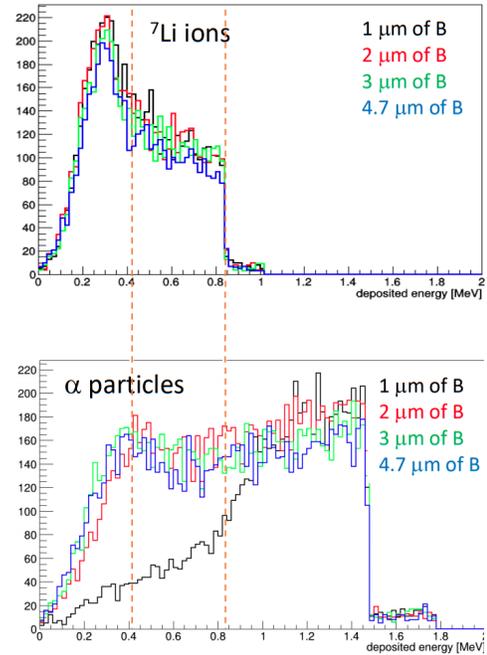

Fig. 2. Energy released for different Boron thickness for $^7Li$ (top) and α (bottom) in four different simulations of 1M thermal neutrons impinging on the Boron layer.

The planar converter shows its maximum neutron conversion between 2 and 3 μm $B_4C$ thickness.

The second design under study is the one with the *grooved cathode*. This design has to maximize the neutron conversion with sharp spikes and a small spacing from one side and the electron extraction from the depth from the other side. The first prototype of the grooved cathode has depths of 2.5 mm, a slope of 10° and a spacing of 0.5 mm. The Boron sputtering thickness of this layer is 2.5 μm. The simulation of the deposited energy as a function of the production region (spacing, slope and peak) is shown in Fig. 3: the differences in the distributions are related to the path length of the particles and to the cathode geometry. Outside the peak regions, the probability to impact the next peak is higher and this reduces the particle path, therefore the deposited energy.

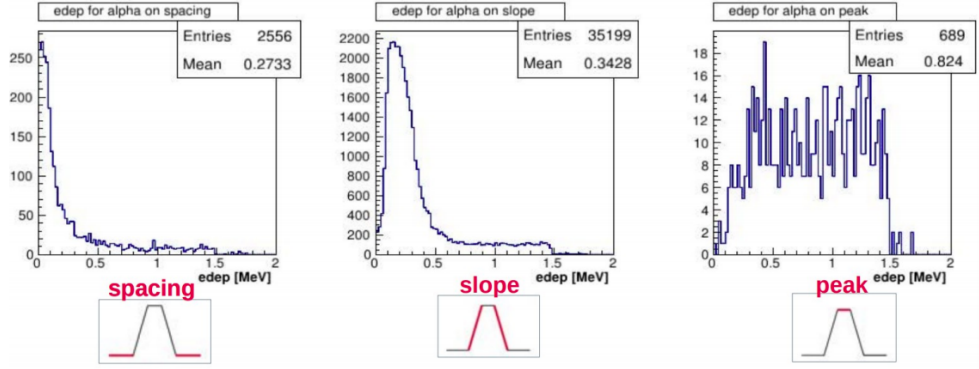

Fig. 3. Deposited energy distribution of α particles produced by 1M thermal neutrons on the grooved cathode for three different regions: spacing, slope and peak. Similar distribution can be found for $^7$Li particles too.

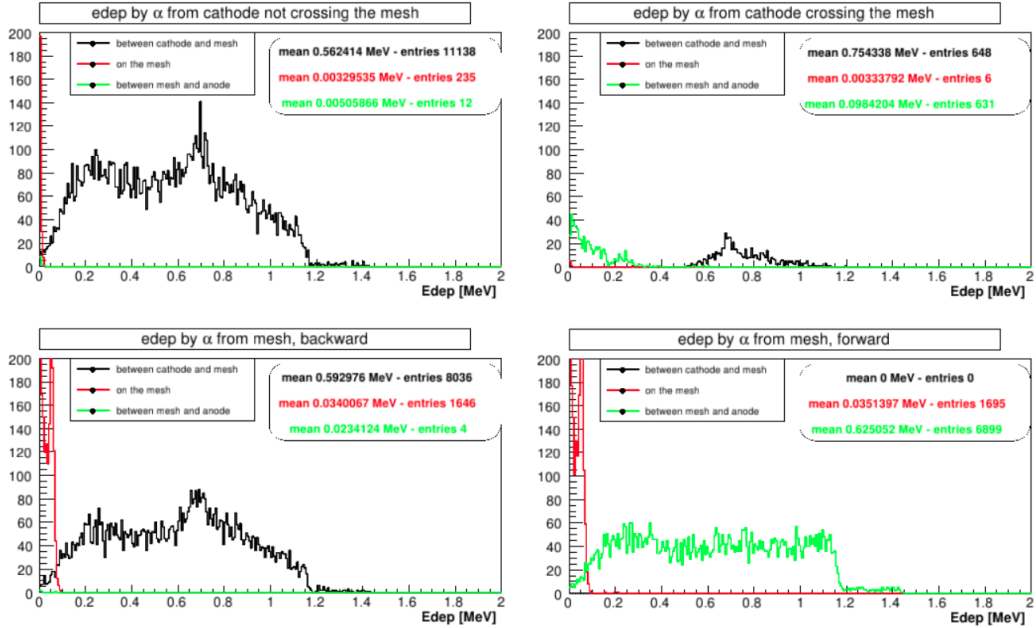

Fig. 4. Deposited energy distributions for α particles produced on a μRWELL having a boron-coated cathode and a boron-coated mesh. The distributions are separated in regions to define the amount of expected primary electrons as a function of the ionization regions. Four different conditions were simulated: α particles from the cathode not crossing the mesh, α particles from the cathode crossing the mesh, α from the mesh to the cathode and α from the mesh to the readout.

The third design under study is the one with *boron-coated mesh*. The one used in the measurements has an optical transparency of 33% and with a proper electrical configuration it is possible to extract the largest number of electrons generated in the ionization. The contribution of the neutron conversion coming from the cathode and the mesh are measured separately, as a function of the α and $^7$Li directions, as shown in Fig. 4.

The simulations described above are used to measure the neutron conversion probability as a function of the Boron thickness and the effects related to the involved geometry (planar, grooved or mesh). This information together with the detector settings will provide the expected performance of the μRWELL in the experimental measurement proposed in the next sections.

V. EXPERIMENTAL SETUP

A preliminary characterization of the prototypes was done at the ENEA HOTNES in Frascati [8]. The source is a calibrated $^{241}$Am-B thermal neutron source placed in a cylindrical cavity delimited by polyethylene walls. HOTNES exploits a polyethylene shadow bar that prevents fast neutrons from directly reaching the samples. The effect of the shadow bar and of the cavity walls combined in such a way is that the thermal neutron fluency is nearly uniform. The resulting fluence rate at the HOTNES reference irradiation plane is

758±16 cm$^{-2}$s$^{-1}$. Experimental measurements and simulations evaluate the neutron fraction absorbed by the mechanical structure of the detector: a fraction of 17.7% is removed from the neutron fluence rate. A representation of the HOTNES setup is shown in Fig. 5.

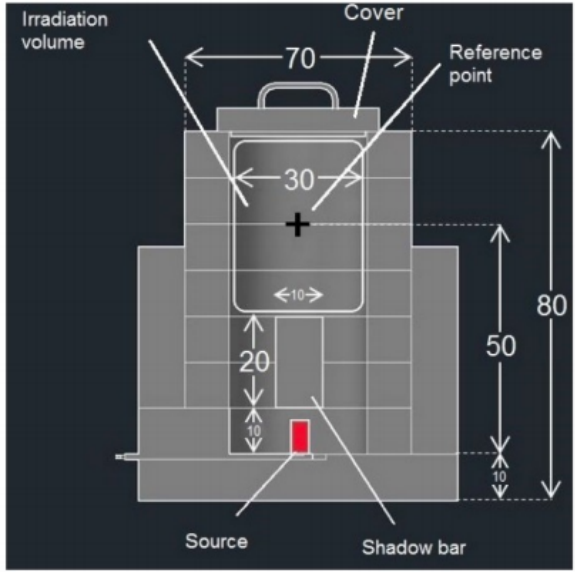

Fig. 5. Representation of the HOTNES setup with the radioactive source and the shadow bar in the middle The polyethylene blocks the neutrons to remove any hazard.

The energy distribution of the neutron generated by HOTNES is not centered around 25 meV but it is slightly higher. The energy peak is around 100 meV and the FWHM is around 290 meV. This has an impact on the experimental and the simulation results because the neutron cross section depends mainly on their energy: higher energy leads to smaller efficiency. A factor two of reduction is expected with the HOTNES setup with respect to the one evaluated with thermal neutrons.

## VI. MEASUREMENT TECHNIQUES

Neutrons release in the gas volume a large number of electrons and a proper amplification can be used to increase it up to several hundreds of fC. The two measures used in this paper exploit the same signal with different techniques: the *current mode* and the *counting mode*. Thanks to an A1561HDM CAEN power supply with a 50-100 pA sensitivity current monitor, it is possible to measure the neutron detection efficiency. The current flowing to the resistive layer of the μRWELL is proportional to the electrons released in the gas and the detector gain as shown in (2).

$$i = \Phi * \varepsilon * N * G * S \quad (2)$$

where i is the current, $\Phi$ is the neutron flux at the reference point, $\varepsilon$ is the neutron conversion efficiency, N is the average number of electrons coming from the ionization (contribution evaluated from simulation), G is the detector gain and S is the detector surface.

The setup with the mesh has four different contributions as shown in Fig. 1 and 4. The proper electron transparency (evaluated from simulation) has to be taken into account as shown in (3).

$$i_1 = \Phi * \varepsilon_1 * N_1 * G * S * T_{elec}$$
$$i_2 = \Phi * \varepsilon_2 * N_2 * G * S$$
$$i_3 = \Phi * \varepsilon_3 * N_3 * G * S$$
$$i_4 = \Phi * \varepsilon_4 * N_4 * G * S * T_{elec} \quad (3)$$

where $T_{elec}$ is the electron transparency and the cases from 1 to 4 are the different production layers inside the detector as shown in Fig. 1 right. A detector gain calibration was made using an X-ray gun, covering the operating range of the measurements (gain from 300 to 700).

The other measurement technique is the counting mode developed with several custom electronic boards based on CAEN A1422 or CREMAT CR-110 to cope with the signal produced by a μRWELL and α or $^7$Li particles, as shown in Table 2. This electronics readout has to read a signal of about hundreds pC from all the strips of the μRWELL. Several tests were performed to optimize the electronics with the detector before the source test by means of cosmic rays and Strontium radioactive source.

| Pre-amplifier chip | CAEN A1422 | CREMAT CR-110 |
|---|---|---|
| gain | 0.2 mV/fC | 0.5 mV/fC |
| noise | 20 mV | 20 mV |
| signal duration | 5 μs | 5 μs |

Table 2. Custom electronics' main features for the counting mode are listed.

Each signal is then acquired with a Picoscope digitizer. The neutron flux is measured directly and the ratio with the nominal values returns the neutron detection efficiency.

## VII. RESULTS

The neutron efficiency is evaluated in the current mode by means of simulation results. This technique is very simple and it does not need any electronics. The current expected from simulation is about a few nA, depending on the gain and the design chosen. The experimental neutron efficiency evaluated with the planar cathode design shows a variation as a function of the Boron thickness. The first measurement, reported shown in Fig. 6, compares the experimental results using (2) with the simulations[5].

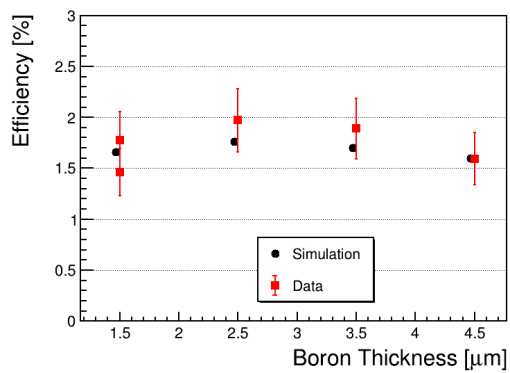

Fig. 6. Neutron detection efficiency of a μRWELL detector with planar boron-coated cathode as a function of the boron thickness is shown.

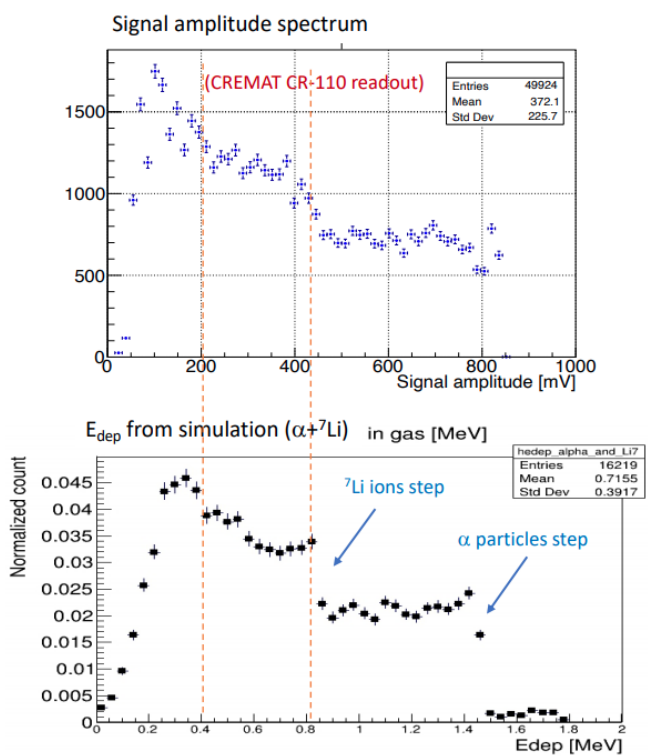

Fig. 7 Signal amplitude spectrum from experimental results (top) and deposited energy from α and $^7$Li together from simulations (bottom) are shown. Dotted lines are drawn to connect similar shapes of the two connected quantities. Both plots have been evaluated with the planar cathode design.

The current mode measurement was applied also to the mesh design, where an efficiency of 4.6 ± 1.0 was measured using (3). Systematic measurements were performed to validate the current mode: a μRWELL without a coated cathode was used to measure the current measurement contribution from the HV. No current contributions above 0.1 nA have been found up to a gain of 1000 both inside and outside HOTNES. A contribution up to 1 nA has been found at a gain of $10^4$ inside HOTNES probably due to the photon's contribution. All the measurements in current mode have been performed with a gain between 300 and 700.

The counting mode measurements were performed for the planar design and the grooved cathode. These measurements are independent from the simulation and any gain calibration. An efficiency of 2.4 ± 0.1 has been measured for the planar cathode and a value of 3.2 ± 0.1 for the grooved cathode. Together with the neutron rate, thanks to the Picoscope it is possible to acquire the induced signal then its amplitude. The signal amplitude distribution with a planar cathode is compared with the one from simulations and the two shapes show similar behaviors, as shown in Fig. 7.
The systematic errors in the counting mode depend on the threshold chosen and the electronic dead-time. A threshold scan has been performed to properly set the threshold above the noise level; while from the electronics side a 20 μs dead time has been chosen to reject any other contribution due to noise correlated to signal contributions.

## VIII. CONCLUSIONS

A neutron detector based on μRWELL technology was designed, manufactured and tested. Coupling a Boron converted to an MPGD makes it possible to develop a low cost and a high precision detector. Radioactive waste monitors and national security portals can be developed based on this technology with the ongoing technological transfer for large area detector construction. In this paper several designs have been studied by means of a neutron source with a neutron conversion efficiency ranging from 2 to 4.6% with the HOTNES neutron energy distribution. A 10% neutron conversion efficiency is expected with a combination of the different designs and thermal neutrons at 25 meV. No contribution comes from the photons' background. The project's next step will be focused on the finalization of the detector design and the electronics for the counting mode.